\documentclass[conference]{IEEEtran}
\IEEEoverridecommandlockouts
\usepackage{amsmath,amssymb,amsfonts,amsthm}
\usepackage{mathrsfs} % for mathscr
\usepackage[normalem]{ulem} % for strikethrough
\usepackage{algorithmic}
\usepackage[dvips]{graphicx}
\usepackage[inline]{asymptote}
\usepackage{textcomp}
\usepackage{verbatim} % for commenting
\usepackage{xcolor}
\usepackage{svg}
\usepackage{cuted}
\usepackage{subcaption}
\usepackage{lipsum}
\usepackage[left=0.65in, right=0.65in, top=0.76in, bottom=1.05in]{geometry}

\begin{document}

\theoremstyle{plain}
\newtheorem{theorem}{Theorem}
\newtheorem{definition}{Definition}
\newtheorem{lemma}{Lemma}
\newtheorem{corollary}{Corollary}
\newtheorem*{remark}{Remark}
\newtheorem{proposition}{Proposition}
\newtheorem{example}{Example}

\title{Time-Frequency Pilot Sequence Design and LoS Delay-Doppler Estimation}

\author{Aadarsh Devanand and Praful D. Mankar

\thanks{A. Devanand and P. D. Mankar are associated with the Signal Processing and Communication Research Centre (SPCRC) at the International Institute of Information Technology, Hyderabad (IIIT-H), India. Email: aadarsh.devanand@research.iiit.ac.in and praful.mankar@iiit.ac.in.}
}

\maketitle

\begin{abstract}
We present a novel framework for line-of-sight (LoS) delay-Doppler (DD) estimation in dense scattering propagation environments.
We present two time-frequency (TF) domain pilot sequences inspired by the Zadoff-Chu sequence that exhibit desirable autocorrelation properties.
Further, we present a twisted convolution-based approach for LoS DD estimation directly from the TF-domain received signal, avoiding an additional TF to DD transformation, which is commonly found in literature.
Numerical results from simulations demonstrate that the proposed framework significantly outperforms traditional single-carrier Zadoff-Chu sequences in both delay and Doppler estimation over a wide range of Rician fading factor and SNR values.
\end{abstract}

\begin{IEEEkeywords}
Time-Frequency, Delay-Doppler, Pilot Sequence, Zadoff-Chu, Twisted Convolution, Autocorrelation Function. 
\end{IEEEkeywords}

%!TEX root = main.tex
\section{Introduction}
\label{sec:introduction}

Delay-Doppler (DD) estimation is a problem that frequently appears in many applications, such as radar systems \cite{wavelet} and orthogonal time-frequency space (OTFS) modulation-based communication systems \cite{dd_pilots}.
Specifically, estimating the time delay and Doppler shift of the line-of-sight (LoS) component is useful for sensing applications, such as user localisation and tracking \cite{disoul}, as well as communication applications, such as directional beamforming.
The DD estimates can be obtained from the ambiguity function (AF), whose structure (i.e., the behaviour of the main and side lobes), and hence the estimation performance, depend on the pilot signal.
Therefore, the task of designing a pilot signal with certain desirable properties, as will be discussed shortly, is of great importance in building robust sensing and communication systems.

\subsection{Related Works}
In \cite{wavelet} and \cite{wbaf}, the wideband cross-ambiguity function (WBCAF) is used to estimate the LoS delay and Doppler shifts in the absence of multipath components (MPCs).
The authors of \cite{wavelet} consider a semi-active radar scenario, where the pilot signal is unknown at the receiver.
The WBCAF of the signals received at two spatially separated sensors is computed to find the relative delay and Doppler shift.
However, this computation is difficult because it requires generating arbitrarily scaled and delayed versions of the received signals.
Hence, the author relies on a wavelet-transform-based method to make this computation tractable.
In \cite{wbaf}, a cooperative scenario is considered where the pilot signal is known at the receiver a priori.
The WBCAF between the pilot and the received signal is computed, followed by locating its peak to estimate the delay and Doppler shifts.
Neither of these works considers the impact of non-line-of-sight (NLoS) components, which exist in most of the propagation scenarios.
These NLoS components lead to inter-symbol and inter-carrier interferences and hence can degrade estimation performance. Thus, the mitigation of these interferences should be considered during pilot signal design. 

In \cite{prior_ch_est}, the design of an optimal pilot sequence is presented based on minimising the Bayesian Cramer-Rao Bound (CRB) in the multipath scenario.
This approach assumes prior knowledge of the channel and noise covariance.
However, this assumption fails in high-mobility scenarios, wherein the channel decorrelates rapidly, thus rendering the prior knowledge less reliable.

Pilot sequence design in the DD domain \cite{otfs_paper} is presented in \cite{dd_pilots} and \cite{dd_ch_est}, to harness the sparsity offered by the DD channel.
The authors of \cite{dd_pilots} consider the problem of DD channel estimation in a multi-antenna OTFS transmission system.
In particular, they designed two-dimensional (2D) pilot sequences based on code-division multiplexing to deal with inter-pilot interference. 
In \cite{dd_ch_est}, a DD domain pilot sequence design is presented that minimises the CRB in high-mobility scenarios. 
This method involves solving a non-convex optimisation problem. 
Both the above works transform the DD domain pilot into the time-frequency (TF) domain via the inverse symplectic Fourier transform.
This transformation can present significant processing overhead compared to placing the pilot directly in the TF domain.
Moreover, these works assume bi-orthogonality of the transmit and receive pulses to represent the input-output relation in the DD domain as a 2D convolution, as given in \cite{otfs_paper}.
However, this assumption may not hold for most practical pulse shaping methods.

In \cite{genetic_alg}, a genetic algorithm is used to find an optimal pilot sequence that minimises the side-lobe level, constrained by the CRB of delay and Doppler estimation.
Although this algorithm can provide a pilot sequence with desirable properties, genetic algorithms in general have certain drawbacks, such as premature convergence to a local optimum and hyperparameter sensitivity.

To summarise, there is scope to design TF-domain pilot sequences for low-complexity channel estimation and/or LoS DD estimation in wideband systems, particularly in the absence of prior information on the channel's DD behaviour.

\subsection{Contributions}
This paper considers LoS delay and Doppler estimation for a wideband system in a multipath fading scenario.
For this purpose, two Zadoff Chu (ZC) sequence-inspired multi-carrier TF domain pilot sequences, called separable ZC and stacked ZC, are presented alongside a robust estimation framework to extract the LoS delay and Doppler from the received signal.
The proposed sequences are numerically demonstrated to provide significantly lower normalised mean squared error (NMSE) as compared to the conventional single-carrier ZC sequence.

\section{System Model and Mathematical Preliminaries}
\label{sec:systemmodel}

In this section, the wireless system under consideration is discussed.
First, a few signal-processing tools are presented.
This is followed by a discussion of the real-time scenario and the channel input-output relationships.

\subsection{Correlation and Ambiguity Functions} \label{subsec: corr_ambgfn_def}
The discrete ambiguity function \cite{bjorke}, defined below, is a signal processing tool used in DD estimation.
\begin{definition}[Discrete Ambiguity Function]
    The discrete cross-ambiguity function (CAF) between time-domain signals \(\mathrm{x}[n]\) and \(\mathrm{y}[n]\) of length \(L\), is given by,
    \begin{equation}
        A_{\mathrm{x},\,\mathrm{y}}[l,\,k] = \frac{1}{L}\sum_{n=0}^L \mathrm{x}[n] \mathrm{y}^*[n+k] e^{-j2\pi l\Delta fn}.
    \end{equation}
\end{definition}
The behaviour of this function is analogous to computing a sequence of correlations, the \(l\)-th of which is the correlation of \(\mathrm{x}[n]\) with a copy of \(\mathrm{y}[n]\) Doppler shifted by \(l\Delta f\).
Hence, the ambiguity function can be replaced with a 2-D correlation function while working with signals in the TF domain, since delays and Doppler shifts correspond to translations along the time and frequency axes, respectively.
The cross-correlation function (CCF) can be obtained from the CAF by evaluating the latter at \(l=0\), i.e., 
\begin{equation} \label{eqn: ac_from_ambg}
    R_{\mathrm{x},\,\mathrm{y}}[k]=A_{\mathrm{x},\,\mathrm{y}}[0,\,k].
\end{equation}

A modified version of the convolution operation appears in the input-output relationships of channels that involve Doppler shifts, called the twisted convolution.

\begin{definition}[Twisted convolution]
    Given signals \(\mathrm{x}[m,\,n]\) and \(\mathrm{y}[m,\,n]\), their twisted convolution \(\mathrm{z}[m,\,n]\) is given by \text{\cite{twisted_conv_invert_paper}}
    \begin{equation}
        \mathrm{z}[m,\,n] = \sum_l \sum_k \mathrm{x}[l,\,k] \mathrm{y}[m-l,\,n-k] e^{j2\pi (m-l)k}.
    \end{equation}
\end{definition}
This operation can be used to capture the frequency-variant nature of the wideband channel.
It is neither an associative nor a commutative operation, and this needs to be taken into consideration while designing the DD estimation procedure in section \ref{subsec: dd_est}.

\subsection{Network Model} \label{subsec: network_model}
Consider an uplink transmission system involving a single-antenna mobile user-equipment (UE) and a stationary single-antenna base station (BS).
%, as shown in Fig. \ref{fig: network_model}.
The position of the UE in \(2\)D space as a function of time is denoted as \((\mathrm{x}(t),\,\mathrm{y}(t))\) and hence the velocity, denoted as \((v_\mathrm{x}(t),\,v_\mathrm{y}(t))\), has the components \(v_\mathrm{x}(t)=\frac{d}{dt}\mathrm{x}(t)\) and \(v_\mathrm{y}(t)=\frac{d}{dt}\mathrm{y}(t)\).
The estimation of the LoS time delay \(\tau_{\mathrm{LoS}}\) and Doppler shift \(\nu_{\mathrm{LoS}}\) of the UE-BS link happens at the BS, which is located at \((\mathrm{x}_b,\,\mathrm{y}_b)\).

\subsection{Signal and Channel Model} \label{subsec: sg_ch_model}
A multi-carrier signal \(X \in \mathbb{C}^{\mathrm{M}\times\mathrm{N}}\), with unit power (\(\|X\|^2 = \sum_{m=0}^{M-1}\sum_{n=0}^{N-1} |X[m,\,n]|^2 = 1\)) is considered, where \(M\) is the number of frequency sub-carriers and \(N\) is the number of time slots.
Each frequency sub-carrier in each time slot contains a baseband complex symbol.

The maximum possible delay in propagation is assumed to be \(\tau_{\mathrm{max}}\)\footnote{There might exist some MPCs that take more time to reach the BS, but it can be safely assumed that any component that has a higher delay than \(\tau_{\mathrm{max}}\) undergoes enough large-scale fading, i.e., path loss and/or shadowing, that it can be ignored.} and the maximum possible Doppler is assumed to be \(\nu_{\mathrm{max}}\)\footnote{If the propagation environment is assumed to be largely stationary, the only Doppler effect will be that corresponding to the velocity of the UE and the angle made by the velocity vector and the line joining the UE and the first reflecting obstacle.\vspace{0.1in}}.
Thus, the channel \(H(\nu,\,\tau)\) can be defined as a function of delay and Doppler with a continuous support \([-\nu_{\mathrm{max}},\,\nu_{\mathrm{max}}]\times[0,\,\tau_{\mathrm{max}}]\).

The signal after modulation, transmission and demodulation is given (ignoring additive noise) by \cite{otfs_paper}  
\begin{equation} \label{eqn: cont_rx_sg}
    \tilde{Y}(f,\,t) = H(\nu,\,\tau) *_\sigma X[m,\,n] *_\sigma A_{g_\mathrm{Tx},\,g_\mathrm{Rx}}(\nu,\,\tau),
\end{equation}
where \(A_{g_\mathrm{Tx},\,g_\mathrm{Rx}}(\nu,\,\tau)\) is the CAF of the transmit pulse \(g_\mathrm{Tx}\) with the receive pulse \(g_\mathrm{Rx}\).
Now, two assumptions are made to simplify the model.
Firstly, it can be assumed that the transmit and receive pulses are chosen such that \(A_{g_\mathrm{Tx},\,g_\mathrm{Rx}}(\nu,\,\tau)\) is invertible under the twisted convolution operation \cite{twisted_conv_invert_paper}.
This is a more relaxed assumption about the ambiguity function of the transmit and receive pulses than the assumption mentioned in \cite{otfs_paper}, that the pulses are biorthogonal. This condition implies the ambiguity function is a Kronecker Delta function in the delay-Doppler space.
Secondly, the DD channel response can be assumed to be non-zero only in the small regions surrounding the points \((m\Delta f,\,nT)\) in the DD space.
This assumption becomes more accurate as the DD grid \(\{(m\Delta f,\,nT):\,n,\,m \in \mathbb{Z}\}\) becomes finer.
Under these assumptions, \eqref{eqn: cont_rx_sg} can be re-written as
\begin{equation} \label{eqn: channel_io_twisted_conv}
    \tilde{Y}[m,\,n] = H[l,\,k] *_\sigma X[m, n],
\end{equation}
where \(l\) and \(k\) are the Doppler and delay indices respectively.

A Rician propagation channel is considered, which can be expressed as
\begin{equation} \label{eqn: overall_ch}
    H = \kappa H_{\mathrm{LoS}} + \sqrt{1-\kappa^2} H_{\mathrm{NLoS}},
\end{equation}
where \(\kappa=\sqrt{\frac{K}{K+1}}\) is a Rician factor representing the strength of the LoS component \cite{goldsmith}.
\(H_{\mathrm{LoS}}\) is a matrix with only one non-zero entry at the index corresponding to the delay and Doppler shift of the LoS component, such that
\begin{equation}
    H_{\mathrm{LoS}}[l,\,k] =
    \begin{cases}
        \mathcal{P}(k\mathrm{T}) &\text{if}~ (k\mathrm{T},\,l\Delta f)=(\tau_{\mathrm{LoS}},\,\nu_{\mathrm{LoS}})\\
        0 & \text{otherwise,}
    \end{cases}
\end{equation}
where \(\mathcal{P}(\tau)\) denotes the power-delay profile \cite{goldsmith} of our channel, which is considered to be exponential in nature, i.e.,
\begin{equation}
    \mathcal{P}(\tau) = 
    \begin{cases}
        0 & \tau < \tau_{\mathrm{LoS}}\\
        e^{-\beta\tau} & \tau \ge \tau_{\mathrm{LoS}}.
    \end{cases}
\end{equation}
Assuming a dense scattering environment, with a large number of scatterers corresponding to each delay and Doppler value, \(H_{\mathrm{NLoS}}\) can be considered to be a Rayleigh fading channel matrix, i.e., each entry is a complex Gaussian distributed value with variance equal to the \(\mathcal{P}(\tau)\) value at the respective delay, such that
\begin{equation}
    H_{\mathrm{NLoS}}[l,\,k] \sim \mathcal{CN}(0, \mathcal{P}(k\mathrm{T})).
\end{equation}

Considering additive white Gaussian noise, the input-output relationship can be expressed as
\begin{equation} \label{eqn: rx_sg_noise}
    Y[m\,,n] = H[l,\,k] \ *_\sigma \ X[m,\,n] + W[m,\,n],
\end{equation}
where \(W[m,\,n] \sim \mathcal{CN}(0, \frac{1}{\mathrm{SNR}})\), and \(\mathrm{SNR}\) is the signal-to-noise ratio.

\section{Pilot Sequences and Delay-Doppler Estimation}
\label{sec:DD_estimation}

Here, a brief introduction to Zadoff-Chu sequences is given, followed by the design of two TF-domain pilot sequences. Then, a framework to estimate the time delay and Doppler shift of the LoS component of the channel is presented.

\begin{figure}[t!]
     \centering
     \vspace{0.16in}
     \includegraphics[trim={0cm 2.2cm 0 3.7cm},clip,width=0.4\textwidth]{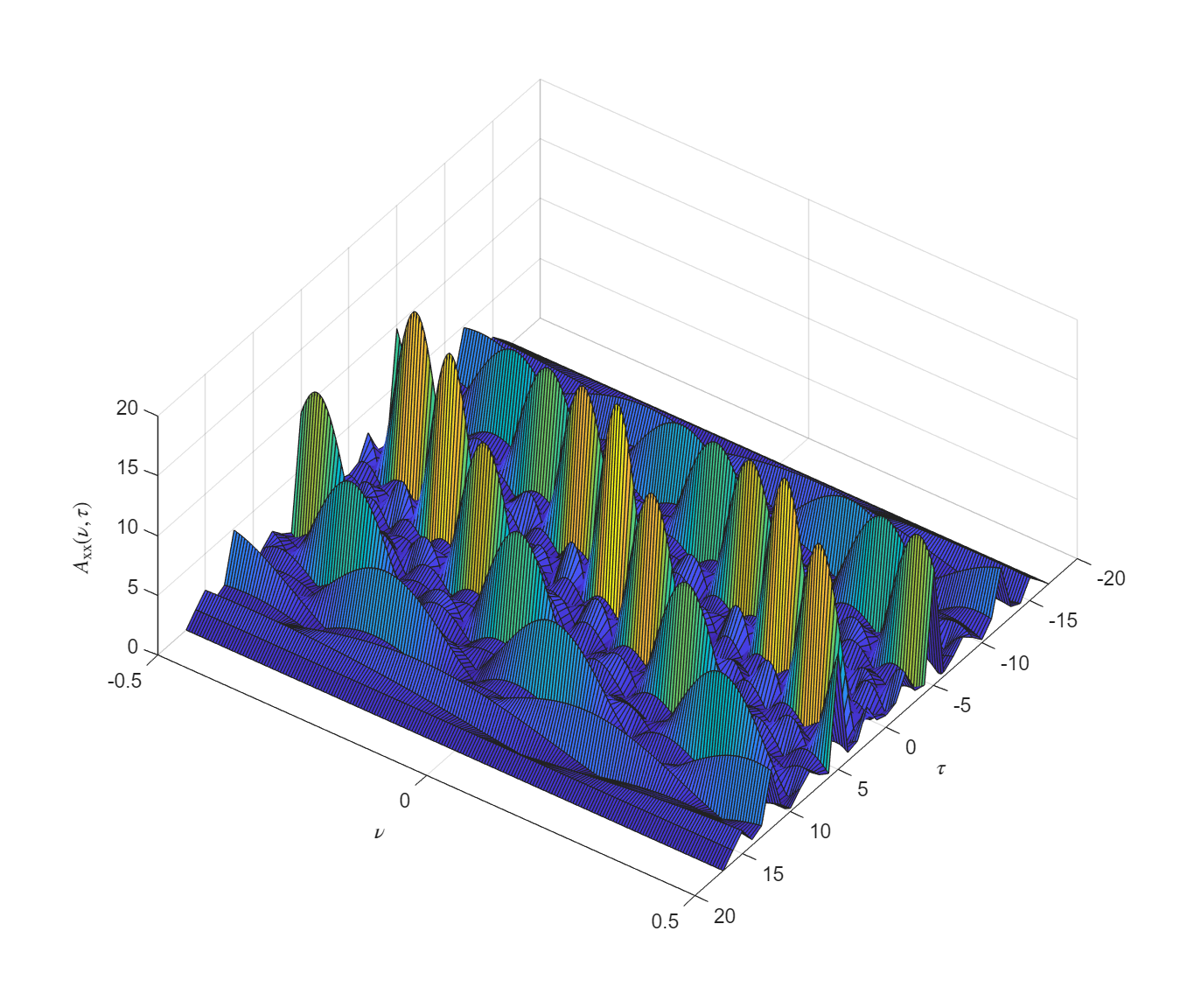}
     \label{fig: zc_ambgfn}
     \caption{\small Ambiguity function of a ZC sequence of  \(L=17\).}
     \vspace{-0.2in}
     \label{fig: zc_autocorr_ambgfn}
\end{figure}

\subsection{Zadoff-Chu Sequence}\label{subsec: ZC_seq}
In delay estimation applications, it is beneficial to have sequences with certain desirable correlation properties.
The periodic correlation of sequences \(\mathrm{x}[n]\) and \(\mathrm{y}[n]\) of length \(L\) is given by \vspace{-0.1in}
\begin{equation}\label{eqn: correlation}
    \bar{R}_{\mathrm{xy}}[k] = \frac{1}{L}\sum_{n=0}^L \mathrm{y}[n] \mathrm{x}^*[(n+k)_L],
    \vspace{0.04in}
\end{equation}
where \((n)_L\) denotes \(n \,\text{mod}\,L\).
The Zadoff-Chu sequence \cite{zadoff_chu} is defined as follows.
\begin{definition}[Zadoff-Chu Sequence]
    A ZC sequence \(s_r[n]\) of length \(L\) and that has root index \(r\) is defined by,
    \begin{equation}
        s_r[n]=\tfrac{1}{L}\exp{\left[-j\pi r\tfrac{n(n+1)}{L}\right]}.
    \end{equation}
    % \vspace{-0.06in}
    The length \(L\) must be an odd number and is usually a prime number.
    The root index \(r\) is an integer co-prime to \(L\).
\end{definition}
The ZC sequence can be proven to have the following two desirable correlation properties.
\begin{enumerate}
    \item \textit{Zero periodic autocorrelation}, i.e., \(\bar{R}_{\mathrm{xx}}[k]=\delta[k]\), and
    \item \textit{Low cross-correlation}, i.e., for any two ZC sequences \(\mathrm{x}\) and \(\mathrm{y}\), with distinct root index values, \(\bar{R}_{\mathrm{xy}}[k]=\frac{1}{\sqrt{L}}\).
\end{enumerate}
Consequently, ZC sequences belong to a family of sequences known as Constant Amplitude Zero Autocorrelation (CAZAC) sequences, which are widely used in delay estimation applications.
Since no cyclic prefix is considered, the channel input-output relationship is represented by a linear/twisted convolution operation instead of a circular convolution operation.
The ambiguity function of this signal has been illustrated in Fig. \ref{fig: zc_autocorr_ambgfn}. 
The linear autocorrelation function (ACF) of the ZC sequence has small side lobes as shown in the figure (at $\nu=0$).
However, the ambiguity function has many significant false peaks.
Thus, the ZC sequence may be useful for delay estimation but not for joint DD estimation.

\begin{figure*}[t!]
    \centering
    % --- Subfigure (a) ---
    \begin{subfigure}[b]{0.24\textwidth}
        \centering
        \includegraphics[trim={0.3cm 0.5cm 0.5cm 1cm},clip,width=\textwidth]{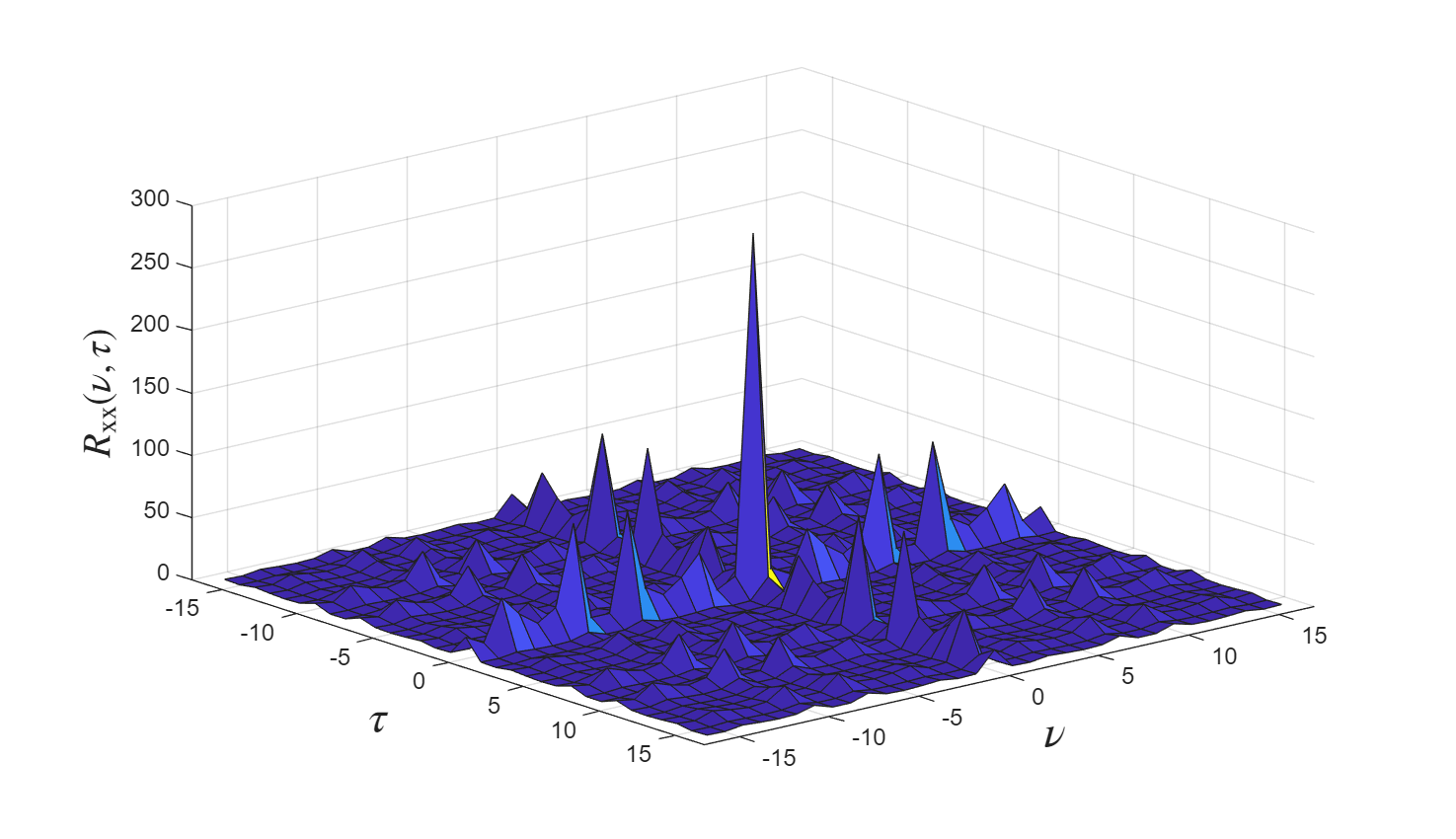}
        \caption{ACF of separable ZC seq.}
        \label{fig:sep_zc_acf}
    \end{subfigure}
    \hfill
    % --- Subfigure (b) ---
    \begin{subfigure}[b]{0.24\textwidth}
        \centering
        \includegraphics[trim={0.3cm 0.5cm 0.5cm 1cm},clip,width=\textwidth]{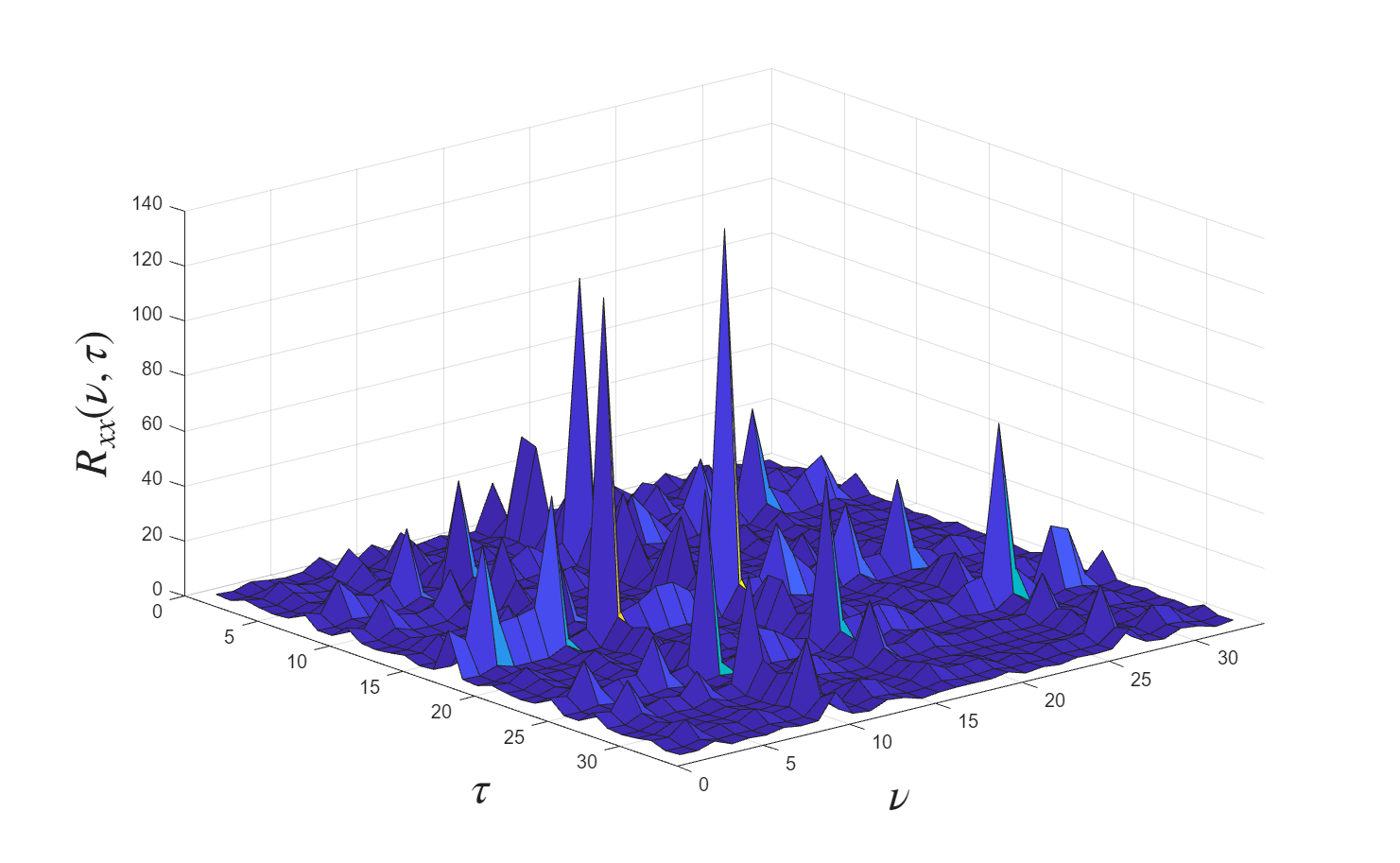}
        \caption{Twisted ACF of separable ZC.}
        \label{fig:sep_zc_twist}
    \end{subfigure}
    \hfill
    % --- Subfigure (c) ---
    \begin{subfigure}[b]{0.24\textwidth}
        \centering
        \includegraphics[trim={0.3cm 0.5cm 0.5cm 1cm},clip,width=\textwidth]{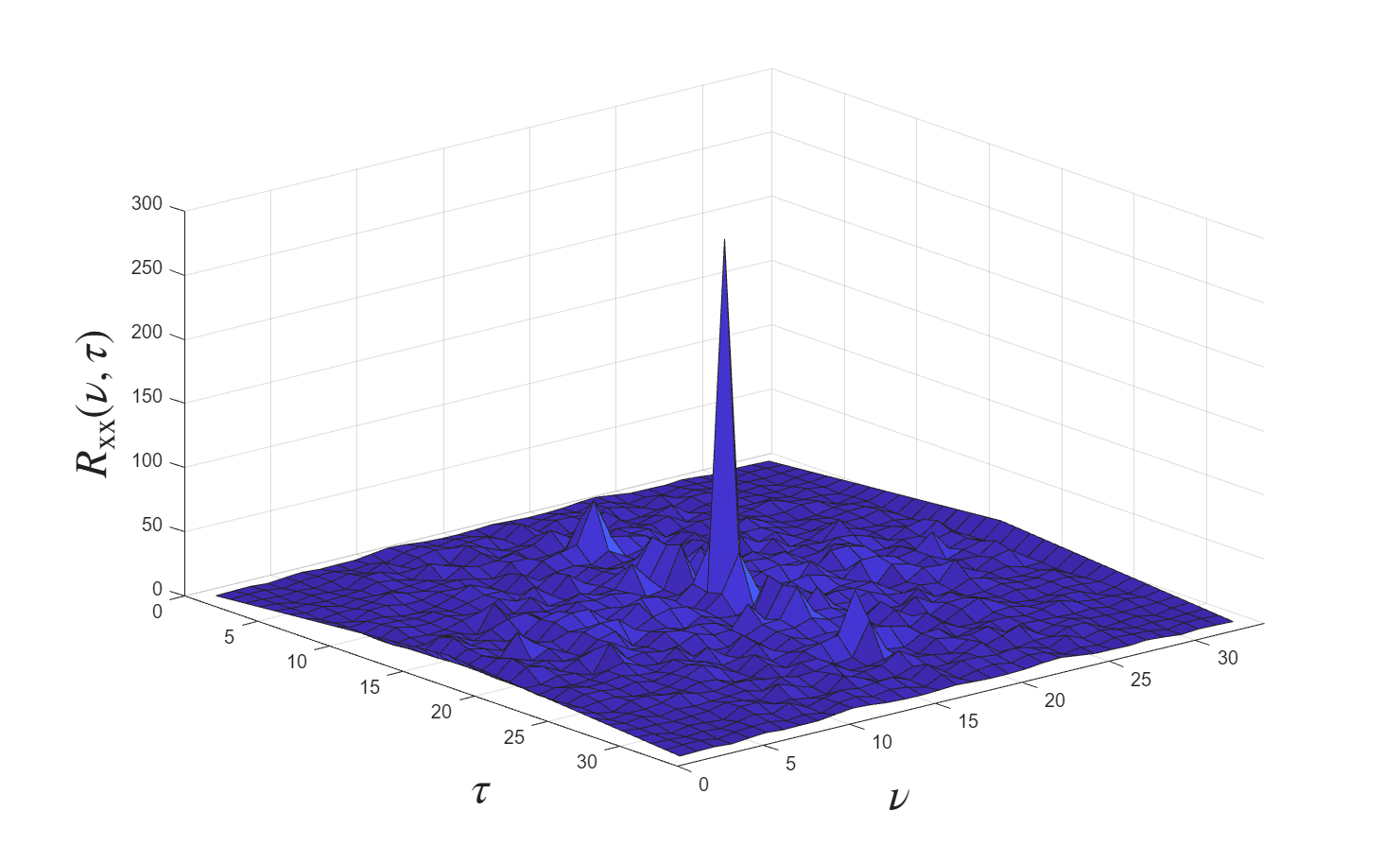}
        \caption{ACF of stacked ZC seq.}
        \label{fig:stack_zc_acf}
    \end{subfigure}
    \hfill
    % --- Subfigure (d) ---
    \begin{subfigure}[b]{0.24\textwidth}
        \centering
        \includegraphics[trim={0.29cm 0.5cm 0.5cm 1cm},clip,width=\textwidth]{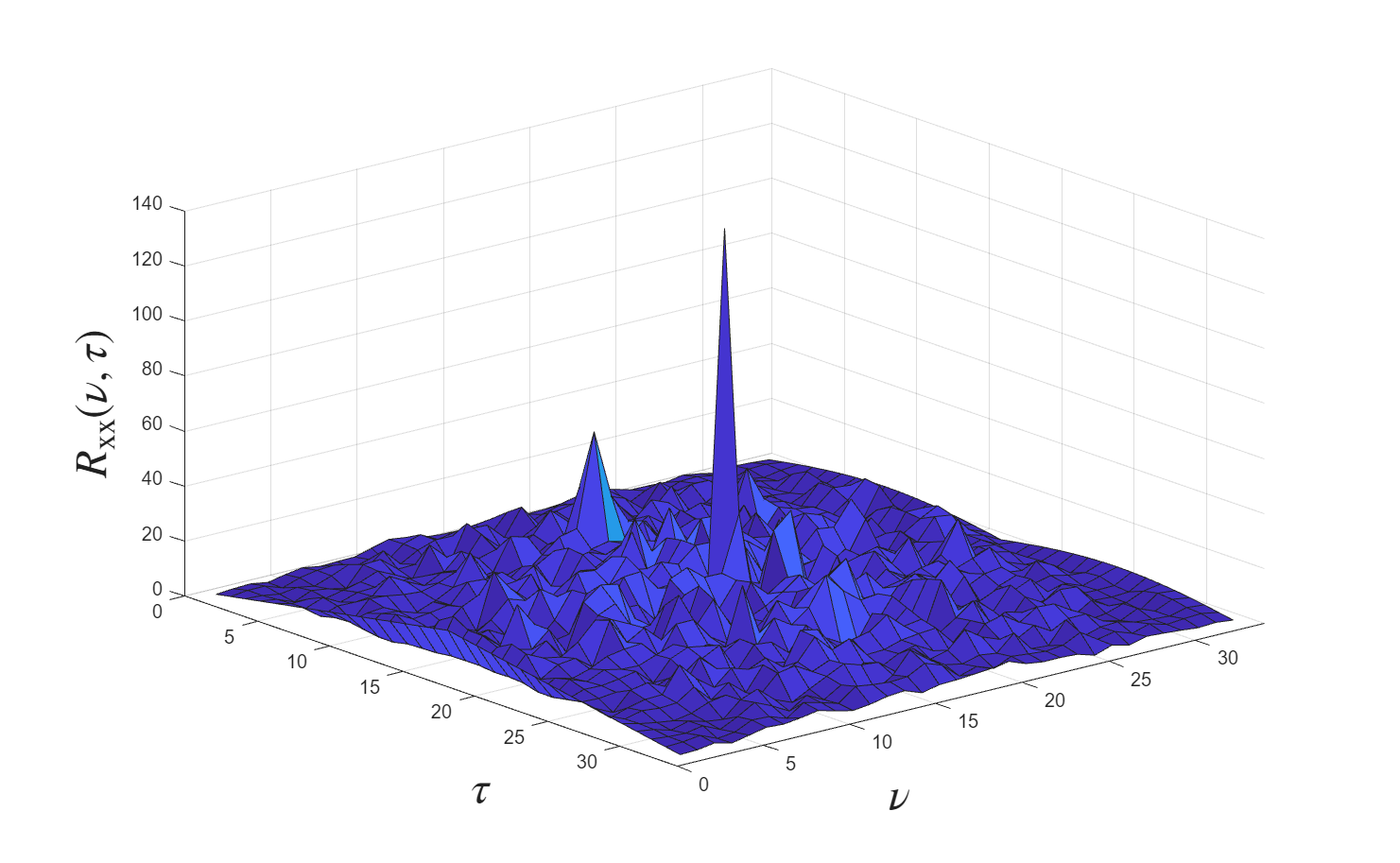}
        \caption{Twisted ACF of stacked ZC.}
        \label{fig:stack_zc_twist}
    \end{subfigure}

    \caption{2D linear and twisted ACFs of separable and stacked ZC sequences for $M=N=17$.}
    \label{fig:2d_zc_autocorrs}
\end{figure*}

\subsection{Time-Frequency Domain Pilots} \label{subsec: tf_zc_seq}
\subsubsection{Separable ZC Sequence}
As mentioned in section \ref{subsec: corr_ambgfn_def}, the ambiguity function is equivalent to a 2D correlation function in the time-frequency domain.
Hence, it would be beneficial to have a sequence with the above correlation properties under both time and frequency shifts.
Since the ZC sequence has desirable correlation properties, it is used to construct the Separable ZC sequence, which is defined as
\begin{equation} \label{eqn: sep_zc_seq}
    F[m,\,n] = \tfrac{1}{MN} \exp{\left[-j\pi \left\{r_f\tfrac{m(m+1)}{M}+r_t\tfrac{n(n+1)}{N}\right\}\right]},
\end{equation}
where \(r_f\) is the root index for the frequency axis and \(r_t\) is the root index for the time axis.
\(M\) and \(N\) are the number of rows and columns, corresponding to the number of frequency sub-carriers and number of time slots, respectively.
The matrix \(F\) can be separated into the outer product of two ZC sequence vectors, one of length \(\mathrm{M}\) and the other of length \(\mathrm{N}\).

The 2D linear and twisted ACFs of the separable ZC sequence are shown in Fig. \ref{fig:sep_zc_acf} and \ref{fig:sep_zc_twist}, respectively. The linear ACF has much less prominent side lobes than the ZC sequence, shown in Fig. \ref{fig: zc_autocorr_ambgfn}.
However, the twisted ACF has some prominent spikes besides that at the origin.
\vspace{0.25cm}
\subsubsection{Stacked ZC Sequence}
A ZC sequence with a unique root index is placed at each frequency sub-carrier, thus the name.
In the construction of this sequence, both the properties of ZC sequences mentioned above are utilised.
The former ensures that the sequence has a desirable autocorrelation on the time axis, and the latter ensures desirable autocorrelation on the frequency axis.
The sequence is defined as
\vspace{0.5cm}
\begin{equation}\label{eqn: stac_zc_seq}
    G[m,\,n] = \tfrac{1}{MN} e^{-j\pi r_m\tfrac{n(n+1)}{N}},
\end{equation}
where \(r_m\) is the root index of the sequence placed at the \(m\)-th frequency sub-carrier.
It is worth noting that the transpose of the above stacked ZC sequence, i.e., a ZC sequence with a unique root index placed along the frequency domain and stacked in the time domain, is a valid time-frequency pilot as well.
The 2D linear and twisted ACFs of the stacked ZC sequence are shown in Fig. \ref{fig:stack_zc_acf} and \ref{fig:stack_zc_twist}, respectively.
Fig. \ref{fig:stack_zc_acf} shows that the ACF of the stacked ZC sequence has almost no side lobes along the \(\tau\) axis (\(\nu=0\)) because it contains sequences with distinct root indices at different subcarriers, which have a cross-correlation of $\tfrac{1}{\sqrt{N}}$.
However, it has small sidelobes along the \(\nu\) axis due to the properties of the ZC sequence. Further, the twisted ACF of the stacked ZC sequence, shown in Fig. \ref{fig:stack_zc_twist}, has much less prominent sidelobes than that of the separable ZC sequence, shown in Fig. \ref{fig:sep_zc_twist}.

\subsection{Delay-Doppler Estimation} \label{subsec: dd_est}
A common method of DD estimation is the matched filter, which in this context implements the CAF of the received signal with the pilot.
In a traditional matched filter, a flipped and conjugated version of the pilot sequence \(X[m,\,n]\) is linearly convolved (1D/2D) with the received signal.
In the case where the channel input-output relationship is represented as the 2D convolution of the channel matrix with the pilot sequence, assuming the used pilot exhibits the desirable correlation properties discussed in Section \ref{subsec: ZC_seq}, the traditional matched filter $X^*[-m,-n]$ provides the estimate of the channel matrix \(H[m,\,n]\), given by
\begin{align*}
    \hat{H}[m,\,n] &= \tilde{Y}[m,\,n] *_2 X^*[-m,\,-n]\\
    & = (H[l,\,k] *_2 X[m,\,n]) *_2 X^*[-m,\,-n]\\
    & \stackrel{(a)}{=} H[l,\,k] *_2 (X[m,\,n] *_2 X^*[-m,\,-n])\\
    & \stackrel{(b)}{\approx} H[l,\,k] *_2 \delta[m,\,n] = H[l,\,k],
\end{align*}
where \(*_2\) denotes 2D convolution, equality (a) follows using the associative property of the linear convolution, and the approximation (b) appears due to the false peaks in the 2D ACF of the ZC sequence (see Fig. \ref{fig:sep_zc_acf} and Fig. \ref{fig:stack_zc_acf}).
However, this approach is not suitable for the received signal model given in \eqref{eqn: rx_sg_noise} as the involved twisted convolution is non-associative.
Thus, an alternative filter \(\Gamma[l,\,k]\) based on twisted convolution is proposed, whose output can be represented as
\begin{equation} \label{eqn: twist_mf}
    {Q}[l,\,k] = \tilde{Y}[m,\,n] *_\sigma \Gamma[l,\,k].
\end{equation}
Setting 
\begin{equation} \label{eqn: twist_mf_ir}
    \Gamma[l,\,k] = X^*[-l,\,-k]e^{j2\pi lk},
    \vspace{0.1in}
\end{equation}
the following can be derived (see Appendix \ref{appendix: twist_mf_eqn_simplification}).
\begin{align} \label{eq: H_hat}
    {Q}[l,\,k] = H[l,\,k] + I[l,\,k],
\end{align}
where $I=I_{\rm sep}$ for the separable ZC sequence and $I = I_{\rm stack}$ for the stacked ZC sequence respectively.

Thus, the estimate of LoS delay and Doppler indices can be obtained as 
\begin{equation}
    (\hat{l},\,\hat{k})=\text{arg } \underset{(l,\,k)}{\text{max }} Q[l,\,k].
\end{equation}
The final LoS delay and Doppler estimates can be obtained as $$(\hat{\nu},\,\hat{\tau})=(\hat{l}\Delta f,\, \hat{k}T).$$

%!TEX root = main.tex
\section{Results and Discussion}
\label{sec: results}

\begin{figure*}[htbp]
    \centering
    \begin{subfigure}{0.425\textwidth}
        \centering
        \includegraphics[width=\textwidth]{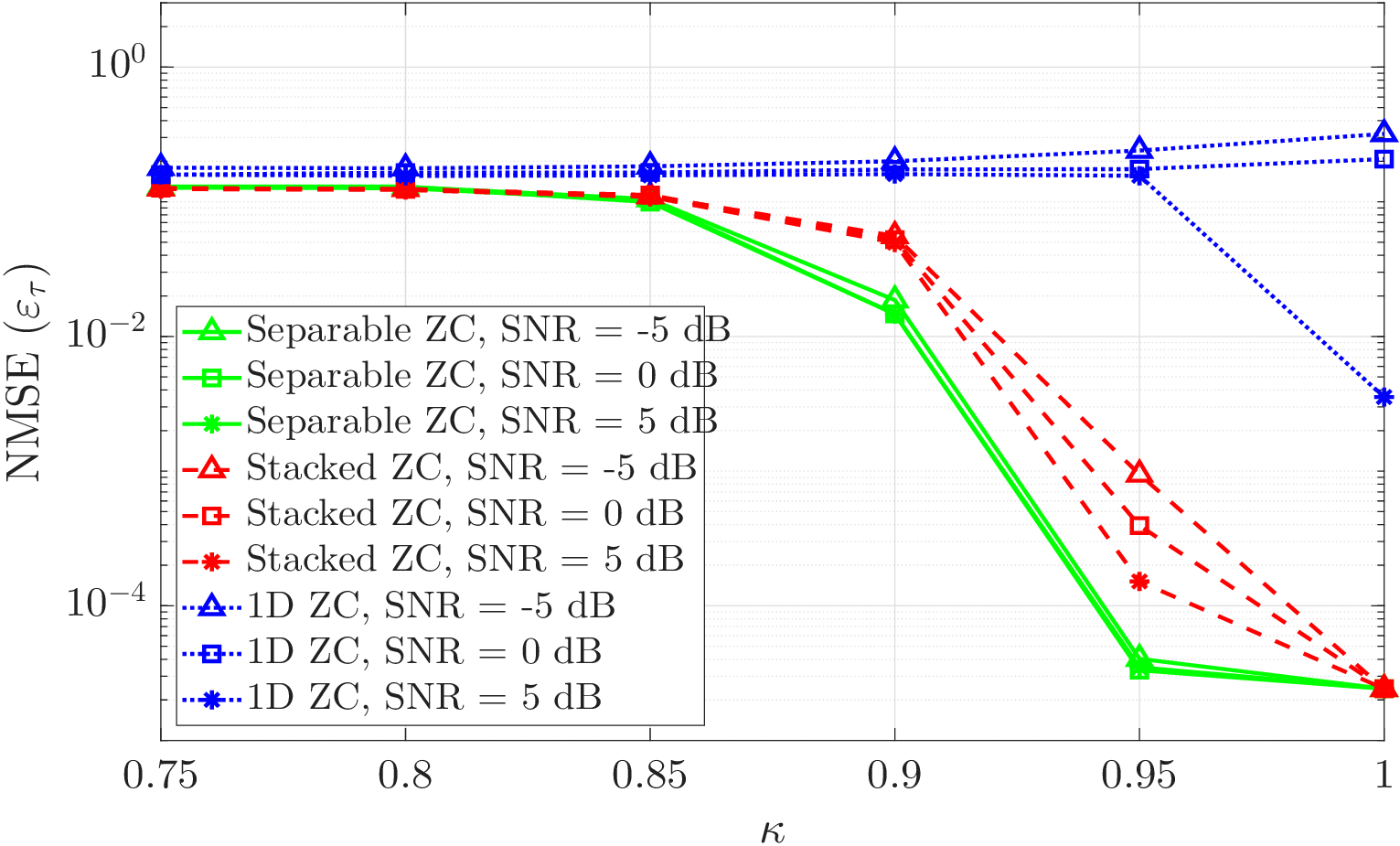}
        \caption{ Time delay estimation.}
        \label{fig: tau_nmse}
    \end{subfigure}
    \hspace{0.25cm}
    \begin{subfigure}{0.425\textwidth}
        \centering
        \includegraphics[width=\textwidth]{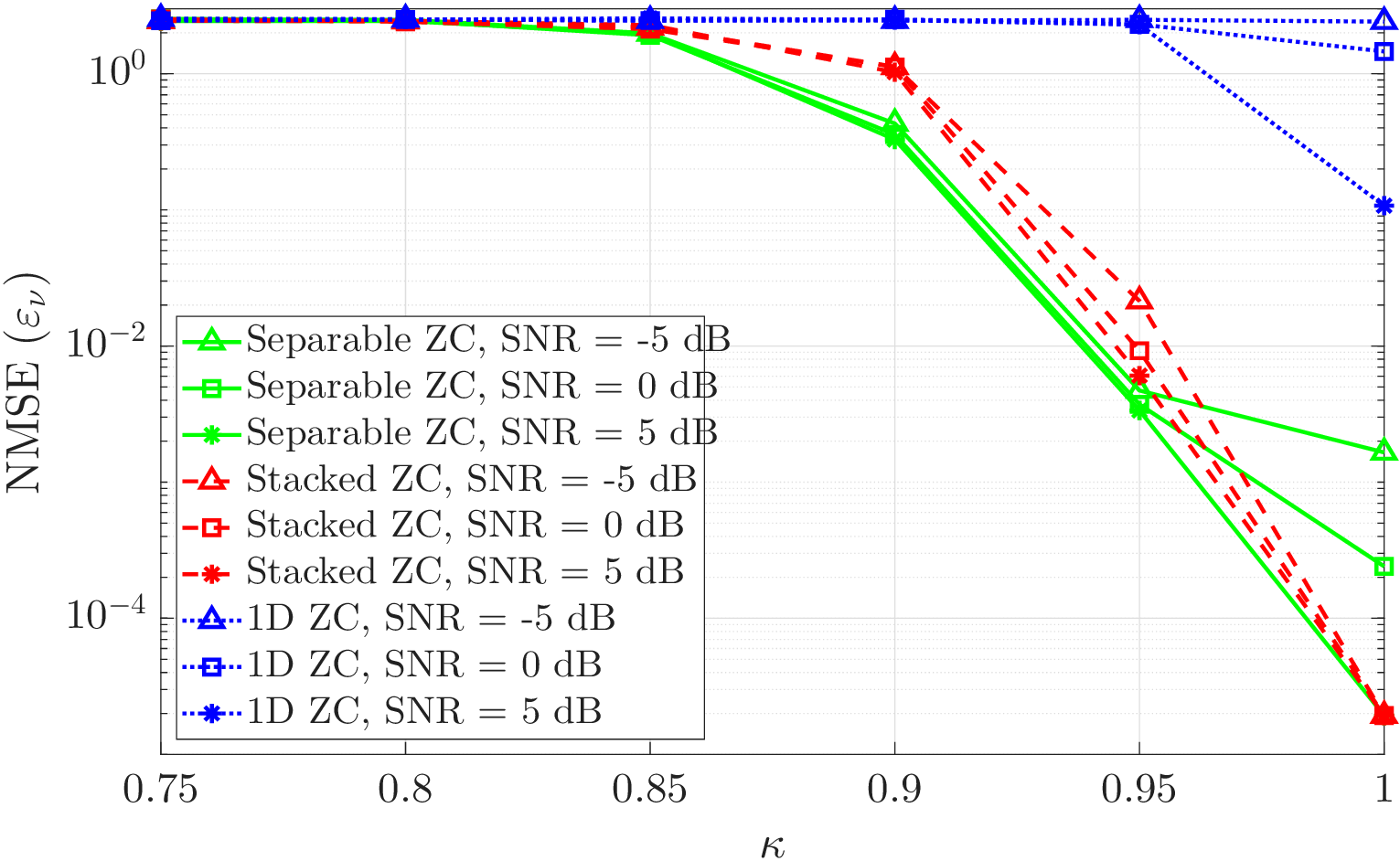}
        \caption{ Doppler shift estimation.}
        \label{fig: nu_nmse}
    \end{subfigure}
    
    \caption{\small Normalised Mean Squared Error (NMSE) for the separable, stacked, and 1D ZC sequences.}
    \label{fig: tau_nu_nmse}
\end{figure*}

This section presents the numerical performance analysis of the LoS DD estimation using the proposed detection method and the ZC sequences  (separable and stacked). The performance of the proposed sequences is compared to that of a single-carrier (1D) ZC sequence, which is widely used in literature (e.g., see \cite{single_ZC}).
The delay and Doppler shift are estimated at regular interval  along the path taken by the mobile UE.
The estimation performance is characterised using NMSE, which is defined as 
\begin{equation} \label{eqn: NMSE_def}
    \varepsilon_\eta = \frac{\sum_{i}|\eta_i-\hat{\eta_i}|^2}{\sum_{i}|\eta_i|^2},
\end{equation}
where \(\eta_i\) and \(\hat{\eta}_i\) denote the true and estimated values of the parameter \(\eta\in\{\tau,\nu\}\) at instant \(i\).
The simulation is performed over the \(10~\mathrm{km} \times 10~\mathrm{km}\) area wherein the BS is considered to be located at  \((\mathrm{x}_b,\,\mathrm{y}_b)=(9.5,\,9.5)~\mathrm{km}\) and the  UE is considered to be moving at speed \(200\) kmph on a circle, whose equation is given by
\begin{align*}
    \mathrm{x}(t) & = 4 + 3.5\cos(0.014~t) \mathrm{km},\\
    \mathrm{y}(t) & = 4 + 3.5\sin(0.014~t) \mathrm{km}.
\end{align*}
The maximum delay and maximum Doppler in this scenario can be obtained as \(\tau_{\mathrm{max}}=50~\mathrm{\mu s}\) and \(\nu_{\mathrm{max}}=1.5~\mathrm{kHz}\), respectively.
To design the DD grid over which the channel response \(H[l,\,k]\) is defined, a time bin size of \(T=0.5~\mu \mathrm{s}\) and frequency bin size of \(\Delta f=10~\mathrm{Hz}\) are considered.
This implies the dimension of \(H[l,\,k]\) will be \(100 \times 150\), with each entry according to \eqref{eqn: overall_ch}.
For the separable and stacked sequences, the dimensions are set as \(M=23\) and \(N=17\).

Fig. \ref{fig: tau_nu_nmse} shows impact of the fading factor $\kappa$ on the NMSEs of estimation of  delay \(\tau\) and Doppler \(\nu\) shifts corresponding to LoS path for \(\text{SNR}\) equal to \(-5,\,0,\) and \(5~\mathrm{dB}\).
It can be observed from the figures that the separable and stacked sequences significantly outperform the 1D ZC sequence in both delay and Doppler estimation, especially when the strength of the LoS component, i.e.  \(\kappa\), is higher.
While the 1D ZC sequence exhibits nearly constant performance across the wide range of \(\kappa\) values, the proposed 2D sequences show improved performance as $\kappa$ increases.
At \(\kappa=1\), both the separable and stacked sequences appear to perform equally well for delay at all SNRs and for Doppler at high SNR.
This implies that while both sequences perform equally well for the LoS channel, separable sequences perform better in the presence of NLoS components, particularly for delay estimation.
The separable ZC sequences perform better than the stacked ZC sequences for Doppler estimation, because along the frequency axis (\(\tau=0\)), the separable ZC sequence has an ideal periodic ACF (\(=\delta(\nu)\)), but the stacked ZC sequence has a periodic ACF equal to \(\frac{1}{\sqrt{M}}\).
It is also apparent from the figure that the estimation performance in both cases improves as SNR increases. Note that the error can be further reduced by increasing the sampling grid resolution.

%!TEX root = main.tex
\section{Summary}
\label{sec:summary}

In this paper, a wideband system was considered wherein the channel input-output relationship was represented as a twisted convolution.
It is worth recalling that this relationship followed from the assumption that the CAF of the transmit and receive pulses is invertible under twisted convolution.
Two TF domain pilot sequences, namely the Separable ZC and Stacked ZC sequences, were presented.
This was followed by a twisted convolution-based approach for LoS DD estimation, whose working, when used with the proposed sequences, was then demonstrated through simulations.
The proposed framework significantly outperforms the single-carrier (1D) ZC sequence in both delay and Doppler estimation in the presence of multipath components.

\vspace{-0.25in}
\begin{strip}
\appendices                                     
\section{Derivation of \eqref{eq: H_hat}} \label{appendix: twist_mf_eqn_simplification}

The output of the filter given in \eqref{eqn: twist_mf} for $\Gamma[l,\,k]$ given in \eqref{eqn: twist_mf_ir} can be simplified as follows
{\allowdisplaybreaks
\begin{align}
    {Q}[l,\,k] & = \tilde{Y}[m,\,n] *_\sigma X^*[-l,\,-k]e^{j2\pi lk}\nonumber\\
    & = \sum_{m'} \sum_{n'} \tilde{Y}[m',\,n']X^*[m'-l,\,n'-k] e^{j2\pi (l-m')(k-n')}e^{j2\pi (l-m')n'} \nonumber\\
    & = \sum_{m'} \sum_{n'} \Big(\sum_{m''} \sum_{n''} H[m'',n''] X[m'-m'',n'-n''] e^{j2\pi(m'-m'')n''} \Big) X^*[m'-l,n'-k]e^{j2\pi(l-m')(k-n')}e^{j2\pi(l-m')n'} \nonumber \\
    & = \sum_{m''} \sum_{n''} H[m'',\,n''] \Big( \sum_{m'} \sum_{n'} X[m'-m'',\,n'-n''] X^*[m'-l,\,n'-k] e^{j2\pi[(m'-m'')n''-(m'-l)k]} \Big) \nonumber\\ \nonumber \\ \nonumber \\
    & = \sum_{m''} \sum_{n''} H[m'',\,n''] e^{j2\pi (lk-m''n'')}\sum_{m'}e^{j2\pi m'(n''-k)}\sum_{n'}X[m'-m'',\,n'-n''] X^*[m'-l,\,n'-k].\label{eq: appendix_A}
\end{align}}
\\
Now, consider the following two cases.\\
\textit{Case I}: When \((m'',\,n'')=(l,\,k)\), the $(l,k)$-th in \eqref{eq: appendix_A} reduces to
\begin{align*}
    H[l,k]\sum_{m'} \sum_{n'} X[m'-l,\,n'-k]X^*[m'-l,\,n'-k] = H[l,k],
\end{align*}
which directly follows from the unit energy of the pilot sequence.\\
\textit{Case II}:  When \((m'',\,n'')\neq(l,\,k)\), \eqref{eq: appendix_A} can be simplified in two different ways, depending on the sequence used.
Substituting the separable ZC sequence, given in \eqref{eqn: sep_zc_seq} (i.e., \(X[m,\,n]=F[m,\,n]\)), all terms except the $(l,k)$-th in \eqref{eq: appendix_A} simplifies to
\begin{align} \label{eqn: sep_interf}
    &I_{\text{sep}}[l,\,k]=\frac{1}{(MN)^2} \underset{(m'',\,n'')\neq(l,\,k)}{\sum_{m''} \sum_{n''}} H[m'',\,n''] e^{j2\pi (lk-m''n'')}\nonumber \\&\hspace{2cm}\left(\sum_{m'}e^{j2\pi m'(n''-k)}e^{-j\pi r_f\frac{(m'-m'')(m'-m''+1)-(m'-l)(m'-l+1)}{M}}\right)\left(\sum_{n'} e^{-j\pi r_t\frac{(n'-n'')(n'-n''+1)-(n'-k)(n'-k+1)}{N}}\right)\nonumber\\
    &~~~= \frac{1}{(MN)^2} \underset{(m'',\,n'')\neq(l,\,k)}{\sum_{m''} \sum_{n''}} H[m'',\,n''] e^{j2\pi (lk-m''n'')} A[m''-l,\,n''-k] R\big[n''-k,\,\max(n'',\,k);N-1-|n''-k|\big],
\end{align}
where \(R\big[n''-k,\,\max(n'',\,k); N-1-|n''-k|\big]\) is the autocorrelation function of the ZC sequence with root index \(r_t\), whose closed-form expression can be found in \cite[Theorem 7]{lin_auto_corr_closedform} as 
\begin{align*}
    R\big[u,\,v;w\big] = \frac{|\sin(\frac{\pi}{N}w(-r_tu)_N)|}{\sin(\frac{\pi}{N}(-r_tu)_N)},
\end{align*}
and \(A[m''-l,\,n''-k]\) denotes the self-ambiguity function of the ZC sequence, whose closed-form expression does not exist.
Therefore, \eqref{eq: appendix_A} for the separable ZC sequence can be expressed as \(Q[l,\,k] = H[l,\,k] + I_{\text{sep}}[l,\,k]\).

In the case of the stacked ZC sequence given in \eqref{eqn: stac_zc_seq} (i.e. \(X[m,\,n]=G[m,\,n]\)), the innermost summation over \(n'\) in \eqref{eq: appendix_A} will be the cross-correlation between two ZC sequences with root indices $r_{m'-m''}$ and $r_{m'-l}$.
Thus,  \eqref{eq: appendix_A} will become 
\begin{align} \label{eqn: stac_interf}
   I_{\text{stack}}[l,\,k] \approx & \frac{1}{\sqrt{N}} \underset{(m'',\,n'')\neq(l,\,k)}{\sum_{m''} \sum_{n''}} H[m'',\,n''] e^{j2\pi (lk-m''n'')}\sum_{m'}e^{j2\pi m'(n''-k)}.
\end{align}
Therefore, \eqref{eq: appendix_A} for the stacked ZC sequence can be expressed as \({Q}[l,\,k] = H[l,\,k] + I_{\text{stack}}[l,\,k]\).
\end{strip}

\bibliographystyle{IEEEtran}
\vspace{-0.25in}
% \bibliography{references}

\end{document}